\documentclass{ws-ijmpa}

\begin{document}

\markboth{K.~Gallmeister, W.~Cassing, T.~Falter, and U.~Mosel}
{Hadron attenuation at HERMES by (pre)hadronic FSI}

%
\catchline{}{}{}{}{}
%

\title{Hadron attenuation at HERMES by (pre)hadronic FSI}

\author{\footnotesize K.~Gallmeister, W.~Cassing, T.~Falter, and U. Mosel}

\address{Institut fuer Theoretische Physik, Universitaet Giessen\\
D-35392 Giessen, Germany}

\maketitle


\begin{abstract}
We investigate hadron formation in high energy electroproduction off complex nuclei in the framework of a BUU transport model. Our approach combines a quantum mechanical treatment of the photon's initial state interactions with a semi-classical coupled channel simulation of the (pre)hadronic final-state interactions (FSI). This allows us to study the hadron attenuation observed at HERMES and to get information on the space-time picture of hadron formation separately for $\pi^{\pm}$, $\pi^0$, $K^\pm$, $p$ and $\bar{p}$.

\end{abstract}

\section{Introduction}	

Hadron formation in deep inelastic scattering off nuclei offers an ideal tool to study the space-time behavior of the hadronization process\cite{Kop}. Depending on whether the struck quark hadronizes inside or outside the nucleus the observed hadron attenuation may be due to quark energy loss via induced gluon radiation\cite{WangArleo} and/or (pre)hadronic final-state interactions FSI\cite{Acc03,Falter,Kop03}.

\section{Model}
Based on the earlier work of Ref.~\refcite{Effe} we have developed in Refs.~\refcite{Falter} and \refcite{Fal02} a method to describe high-energy photon and electron induced reactions on complex nuclei in the framework of a semi-classical BUU transport model. 

In the kinematic regime of the HERMES experiment the virtual photon does not always couple directly to the quark inside the nucleon. Depending on its energy $\nu$ and virtuality $Q^2$ it may also fluctuate into a vector meson or perturbatively branch into a quark-antiquark pair that subsequently scatters off the target. Its elementary reaction with a bound nucleon is modeled using the event generator PYTHIA\cite{PYTHIA}. Thereby we take into account nuclear effects like shadowing, Fermi motion, Pauli blocking and nuclear binding.

The interaction of the photon with the nucleon leads to the excitation of hadronic strings that fragment into hadrons according to the Lund mechanism\cite{Lund}. The Lund model involves three timescales for the production of a hadron: The production proper time $\tau_{p1}$ of the hadron's first constituent, the production proper time $\tau_{p2}$ of the second constituent, and the formation proper time $\tau_f$ of the hadron. In the Lund model the latter corresponds to the space-time point where the world lines of the two hadron constituents cross for the first time. The production time $\tau_{p2}$ can be interpreted as the production time of a color neutral prehadron. For each hadron in a fragmentation process we extract these three proper times directly from the JETSET fragmentation routines which are implemented in PYTHIA.

The reaction products are then propagated in the coupled channel transport model which allows for a probabilistic treatment of the FSI beyond simple absorption mechanisms. The elastic and inelastic FSI of the (pre)hadrons with the nucleons inside the nucleus lead to the production of new particles and to a redistribution of energy and momentum.

\section{Results}
\begin{figure}
\centerline{\psfig{file=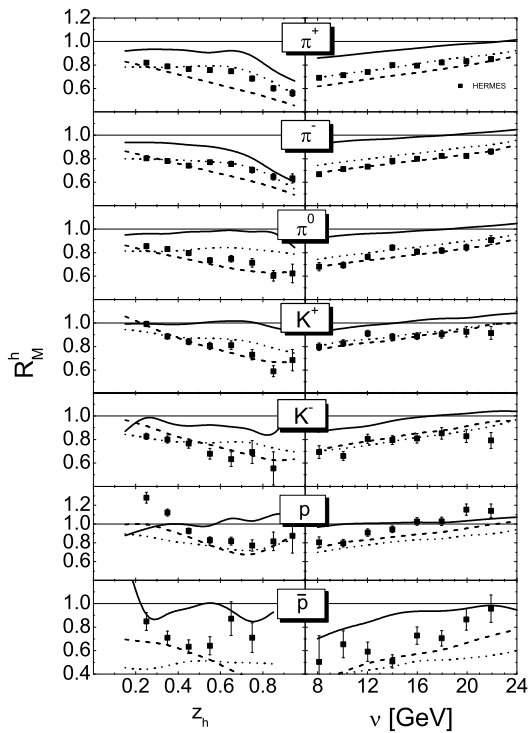,width=7cm}}
\vspace*{8pt}
\caption{Multiplicity ratios of $\pi^{\pm}$, $\pi^0$, $K^\pm$, $p$ and $\bar{p}$ for a $^{84}$Kr nucleus as a function of the hadron energy fraction $z_h$ and the photon energy $\nu$. In the simulation we use the proper times $\tau_{p_2}$ (solid line), $0.2\tau_{p_2}$ (dotted line) and $\tau_{p_1}$ (dashed line) from the JETSET routines as the prehadron production time. The prehadronic cross section is set to the full hadronic cross section and interactions before the production time are neglected.}\label{fig:KrJETSET}
\end{figure}

In Fig.~\ref{fig:KrJETSET} we show our results for the multiplicity ratio
\begin{equation}
R_M^h(z_h,\nu)=\left(\frac{N_h(z_h,\nu)}
{N_e(\nu)}\right)_A\bigg/\left(\frac{N_h(z_h,\nu)}{N_e(\nu)}\right)_D
\end{equation}
for $\pi^{\pm}$, $\pi^0$, $K^\pm$, $p$ and $\bar{p}$ for a $^{84}$Kr nucleus in comparison with the experimental HERMES data\cite{HERMESDIS}. Here $N_h$ denotes the number of hadrons with fractional energy $z_h=E_h/\nu$ and $N_e$ is the number of deep inelastically scattered positrons. For simplicity we set the cross sections of the prehadrons to their hadronic values. Therefore $R_M^h$ is independent of the hadron formation time $\tau_f$ in our calculations.

The solid line shows the result of a calculation where we used the Lund time $\tau_{p2}$ as the production proper time of the prehadrons. As one can see, this yields a too weak hadron attenuation. One has to reduce $\tau_{p2}$ by a factor of 0.2 (dotted line) to get a reasonable agreement with experimental data. Note that a satisfying agreement with experimental data is also achieved by using the Lund time $\tau_{p1}$ as the prehadron production time (dashed line).

\section{Conclusion}
According to the Lund model the dramatic reduction of $\tau_{p2}$ has to be interpreted as an {\it increase} of the string tension $\kappa$ by an unreasonably large factor of about {\it five} in the nuclear medium. By using $\tau_{p1}$ as the starting time of the FSI we account for the interaction of the nucleon debris with the nuclear medium right after the moment of the $\gamma^*N$ interaction. This effectively accounts for the interactions of the hadronic string that is produced in the DIS and that may interact with a hadronic cross section right from the beginning\cite{Cio02}. For a more extensive investigation of hadron attenuation in DIS we refer the reader to Ref.~\refcite{Falter}.

\end{document}